\documentstyle[12pt]{article}
\def\singlespace {\smallskipamount=3.75pt plus1pt minus1pt
                  \medskipamount=7.5pt plus2pt minus2pt
                  \bigskipamount=15pt plus4pt minus4pt
                  \normalbaselineskip=15pt plus0pt minus0pt
                  \normallineskip=1pt
                  \normallineskiplimit=0pt
                  \jot=3.75pt
                  {\def\smallskip {\vskip\smallskipamount}}
                  {\def\medskip   {\vskip\medskipamount}}
                  {\def\bigskip   {\vskip\bigskipamount}}
                  {\setbox\strutbox=\hbox{\vrule 
                    height10.5pt depth4.5pt width 0pt}}
                  \parskip 7.5pt
                  \normalbaselines}
\def\middlespace {\smallskipamount=5.625pt plus1.5pt minus1.5pt
                  \medskipamount=11.25pt plus3pt minus3pt
                  \bigskipamount=22.5pt plus6pt minus6pt
                  \normalbaselineskip=22.5pt plus0pt minus0pt
                  \normallineskip=1pt
                  \normallineskiplimit=0pt
                  \jot=5.625pt
                  {\def\smallskip {\vskip\smallskipamount}}
                  {\def\medskip   {\vskip\medskipamount}}
                  {\def\bigskip   {\vskip\bigskipamount}}
                  {\setbox\strutbox=\hbox{\vrule 
                    height15.75pt depth6.75pt width 0pt}}
                  \parskip 11.25pt
                  \normalbaselines}
\def\doublespace {\smallskipamount=7.5pt plus2pt minus2pt
                  \medskipamount=15pt plus4pt minus4pt
                  \bigskipamount=30pt plus8pt minus8pt
                  \normalbaselineskip=30pt plus0pt minus0pt
                  \normallineskip=2pt
                  \normallineskiplimit=0pt
                  \jot=7.5pt
                  {\def\smallskip {\vskip\smallskipamount}}
                  {\def\medskip   {\vskip\medskipamount}}
                  {\def\bigskip   {\vskip\bigskipamount}}
                  {\setbox\strutbox=\hbox{\vrule 
                    height21.0pt depth9.0pt width 0pt}}
                  \parskip 15.0pt
                  \normalbaselines}

\begin{document}

\author{T. P. Singh,  \\ 
Department of Astronomy and Astrophysics,\\Tata Institute of Fundamental
Research,\\Homi Bhabha Road, Mumbai 400 005, India.}
\title{A characterization of the central shell-focusing singularity
in spherical gravitational collapse}
\date{}
\maketitle

\begin{abstract}
\noindent We give a characterization of the central shell-focusing curvature
singularity that can form in the spherical gravitational collapse of a
bounded matter distribution obeying the dominant energy condition. This
characterization is based on the limiting behaviour of the mass function in the
neighbourhood of the singularity. Depending on the rate of growth of the mass
as a function of the area radius $R$, the singularity may be either covered or 
naked. The singularity is naked if this growth rate is slower than $R$,
covered if it is faster than $R$, and either naked or covered if the growth
rate is same as $R$.
\end{abstract}

\middlespace

\section{Introduction}

\noindent An outstanding open problem in classical general relativity and
relativistic astrophysics is to determine the final fate of the continual
gravitational collapse of a star. The singularity theorems show that, given
a few physically reasonable assumptions, such a collapse will terminate in a
gravitational singularity. However, the theorems do not by themselves imply
that the star ultimately becomes a black hole. In order for a black hole to
form, the singularity must be invisible to far away observers. The theorems
however allow for the possibility that the singularity may be naked, i.e.
visible to a far away observer. Naked singularities, if they were to form in
gravitational collapse, are expected to have theoretical and observational
properties very different from those of black holes. Because of the
difficulty in obtaining a general solution of Einstein equations, it is not
known whether classical general relativity generically admits the formation
of naked singularities in gravitational collapse \cite{pen}.

Over the years, various examples of the formation of covered (i.e. not
naked) and naked singularities have been found in general relativity,
largely in studies of spherical gravitational collapse. Typically, it has
been observed that for any given equation of state, both naked and covered
singularities arise in these examples, depending on the choice of the
initial density and velocity distribution of the star. While on the one hand
there is a similarity amongst these various examples, suggesting an
underlying pattern, on the other hand the issue of the genericity of these 
examples is an open one.

In the present paper we give a characterization of covered and naked
singularities in spherical collapse, on the basis of the behaviour of the
mass function in the neighbourhood of the singularity.

Our characterization is related to an elementary observation regarding
light propagation in Newtonian gravity. The divergence of the density at the
center of a collapsing spherical Newtonian star can be taken to mean that the 
ratio $M/R^3$
diverges as $R\rightarrow 0$, where $M(R)$ is the mass to the interior of
the radius $R$. However, according to the usual escape velocity argument,
the escaping of light (moving at speed $c$) from this density singularity is
governed by the ratio $2M(R)/R$, not by the ratio $M/R^3$. If in the
approach to the singularity, $2M(R)/R$ goes to a limiting value less than
unity, light will be able to escape, but not otherwise. We may write 
\begin{equation}
\label{eff}\frac{2M}R=\frac{2M}{R^3}\;R^2 
\end{equation}
and it is evident that if $M/R^3$ diverges then $2M/R$ may either diverge or
go to zero or be finite and non-zero, depending on whether $M/R^3$ diverges
faster than $1/R^2$, slower than $1/R^2$, or as fast as $1/R^2$,
respectively. On the other hand, if $2M/R$ diverges, then $M/R^3$ also
diverges.

Continuing with this Newtonian argument, let us consider the dependence of
the mass function $M(R)$ on the radius $R$, near the center, at any fixed 
time $t$. Clearly, $M(R)$ will grow slower than $R^{0}$, and faster than or 
equal to $R^{3}$, if we assume that the density at the center is
non-zero, and if $M(R=0)=0$. If the evolution is non-singular, then $M$ goes 
as $R^{3}$. A density singularity will form if there comes a stage during the 
evolution when $M$ goes faster than $R^{3}$. We
know on the basis of theoretical and observational evidence that for a typical
equation of state there are initial conditions (distribution of density and 
velocity) for which the collapsing matter cloud does not become singular, and 
either rebounces or attains equilibrium. These initial conditions correspond 
to selecting $M$ behaving as $R^{3}$ throughout the evolution. Let us
now imagine choosing different classes of initial conditions, for the same
equation of state, so as to move away from such non-singular evolutions and
towards possible singular evolutions. In so far as $M(R)$ is concerned, this is
equivalent to changing the final behavior of $M(R)$ from $M$ 
going as $R^{3}$ to $M$ going as faster than $R^{3}$, and approaching $R^{0}$, 
while possibly realizing growth rates between $R^{3}$ and $R^{0}$.
So long as we are considering the class of solutions for
which the final growth rate is between $R^{3}$ and $R$, the ratio $2M/R$ will
go to zero and light will be able to escape the density singularity - this
is an analog of the naked singularity in general relativity. If the growth 
rate lies between $R$ and $R^{0}$, the ratio $2M/R$ will go to infinity and 
light will not be able to escape from the density singularity. This is an 
analog of the covered singularity. If $M$ grows as $R$, either case may be 
realized, depending on the actual value the ratio takes.

Again on the basis of theoretical and observational evidence, we expect that
the black hole type (i.e. covered) singular solutions, which correspond to 
$M$ going as faster than $R$, will be realized in
gravitational collapse, from some initial conditions. As the initial 
conditions for a given equation of state are
varied from the `weak' to the `strong', one goes from the non-singular
class $M\sim R^{3}$ to the black hole type singular class ($M$ grows faster
than $R$). In doing so, one may or may not encounter solutions of the 
intermediate `naked singularity' class
($M$ goes slower than $R$ but faster than $R^{3}$). In this sense
naked singularities, if they occur, lie between non-singular solutions and 
covered singular solutions.

Considering the similarity of spherical Newtonian collapse to the
corresponding general relativistic case, one expects to be able to formulate
such a characterization in spherical general relativity. We show below
how this can be done, by considering the growth of a curvature invariant
like the Kretschmann scalar, and by assuming the dominant energy condition for
the collapsing matter distribution. According to the dominant energy condition,
pressure components cannot exceed the value of the density at any point in
the spacetime.

We also point out that we deal here only with the characterization of
{\it local} nakedness of singularities, i.e. the issue of whether 
outgoing light rays at all terminate on the singularities in the past. We do 
not concern ourselves with the question of whether or not these light rays 
escape to infinity (i.e. whether or not the singularities are globally naked).

\section{A classification of shell-focusing singularities}

We will assume the matter fields to have an energy-momentum tensor of Type I 
\cite{HE}. With respect to an orthonormal basis $(E_0,E_1,E_2,E_3)$, and with 
$E_0$ being timelike, the energy-momentum tensor can be expressed in the form 
$T_{ik}=diag(\rho, p_1, p_2, p_3)$. All known matter fields, including those of
zero rest mass, have an energy-momentum tensor of this form, with the 
exception of directed radiation. $\rho$ is the energy density as measured by an
observer whose world line has a unit tangent vector $E_0$, while the components
$p_{\alpha}$ are the principal pressures in the three spacelike directions.
The dominant energy condition implies that $\rho\geq 0, |p_\alpha|\leq\rho$.

In comoving coordinates $(t,r,\theta ,\phi )$ the spherically symmetric line
element in general relativity is given by 
\begin{equation}
\label{ele}ds^2=e^\sigma dt^2-e^\omega dr^2-R^2d\Omega ^2 
\end{equation}
where $\sigma ,\omega $ and $R$ are functions of $t$ and $r$. The components 
of the energy-momentum tensor of a collapsing matter distribution are given,
in these coordinates, by $T_k^i=diag(\rho ,-p_r,-p_T,-p_T).$ The functions 
$\rho ,\,p_r$ and $p_T$ have the interpretation of being the density, radial 
pressure and tangential pressure of the matter field, respectively.  

If the matter content of the collapsing object is being described as a fluid,
then one must supplement the above description of the energy-momentum tensor
with two equations of state, one each for the radial and the tangential
pressure. If on the other hand the matter content is a fundamental field, say
a massless scalar field, then the components $\rho, p_r$ and $p_T$ of $T_k^i$
follow from the Lagrangian density of the scalar field. 

The Einstein field equations for this system are

\begin{equation}
\label{mprime}\rho =\frac{F^{\prime }}{R^2R^{\prime }}, 
\end{equation}

\begin{equation}
\label{mdot}p_r=-\frac{\dot F}{R^2\dot R}\,, 
\end{equation}

\begin{equation}
\label{sigpri}\sigma ^{\prime }=-\frac{2p_r^{\prime }}{\rho +p_r}+\frac{%
4R^{\prime }}{R(\rho +p_r)}(p_T-p_r), 
\end{equation}

\begin{equation}
\label{omedot}\dot \omega =-\frac{2\dot \rho }{\rho +p_r}-\frac{4\dot R(\rho
+p_T)}{R(\rho +p_r)}, 
\end{equation}
and

\begin{equation}
\label{energy}e^{-\sigma }\dot R^2=\frac FR+f. 
\end{equation}
Here, prime and dot denote derivatives w.r.t. $r$ and $t$, respectively. In
equations (\ref{mprime}) and (\ref{mdot}) we have set $8\pi G/c^4=1$. The
function $F(t,r)$ results from the integration of Einstein equations, and
has the interpretation of being twice the mass to the interior of the shell
labeled by the coordinate $r$. The function $f(t,r)$ is defined by the
relation $e^\omega =R^{\prime \,2}/(1+f)$ and satisfies the condition $f\geq
-1$.

We assume that at time $t=0$ the spherical object starts undergoing
gravitational collapse, and we choose the initial scaling $R(0,r)=r.$ Next,
we assume that a shell-focusing curvature singularity, given by $%
R(t_s(r),r)=0,$ forms during the evolution. At this singularity, the
Kretschmann scalar $R_{abcd}R^{abcd}$ diverges. The shell labeled by the
comoving coordinate $r$ becomes singular at the time $t_s(r)$. In
particular, the `central' curvature singularity, i.e. the one at $r=0$,
forms at the time $t_s(0).$ It is the central singularity that is of primary
interest from the point of view of nakedness, and is the object of study in
the present paper. We will assume that $t_s(0)<t_s(r\neq 0)$ and that the
central shell-focusing singularity is not preceded by any shell-crossing
singularity. Since $F(t,r)$ is twice the mass, it is physically reasonable
to take $F(t,0)=0$, prior to singularity formation. Furthermore, since the
center is the first point to become singular, according to time $t$, we
assume that $F(t_s(0),0)=0$.

Let us consider the propagation of an outgoing radial null geodesic
congruence $\zeta ^i$. As is known, the occurrence of a covered or locally
naked singularity is
determined by the behaviour of the geodesic expansion $\theta =\zeta
_{\,;i}^i$ in the approach to the singularity. If the expansion of the
outgoing null congruence continues to remain non-negative in the approach to 
the singularity, the singularity will be naked. On the other hand, the 
singularity will be covered if $\theta$ becomes negative in the 
approach to the singularity. It can be shown \cite{sin} that
in a spherically symmetric spacetime $\theta $ can be written, using the
form (\ref{ele}) of the metric, as 
\begin{equation}
\label{thet}\theta =\frac{2R^{\prime }}R\left( 1-\sqrt{\frac{f+F/R}{1+f}}%
\right) \zeta ^r. 
\end{equation}

Here, $R^{\prime }$ and $\zeta ^r$ are non-negative quantities. We will assume
that the ratio $2R'\zeta^r/R$ remains non-zero in the limit. While this is an
untested assumption, it is seen to hold at least in those cases when the
dependence of $R(t,r)$ on $r$ is a power law - since that gives 
$R'/R \sim 1/r$, and also, $\zeta^r=dr/dk\sim r/k$. It then follows 
that the necessary and sufficient condition for the singularity to be naked is
that $F/R$ approach a value less than or equal to unity along the outgoing
geodesic $\zeta ^i,$ as it meets the singularity in the past. 
(The only exceptions to this condition arise when $f$ takes the value $-1$ or 
$+\infty$ in the limit. In the former case, the singularity could be covered
even if $F/R$ goes to unity in the limit, and in the latter case it could be 
naked even if $F/R$ takes a limiting value greater than unity.)

The above condition on $F/R$ is nothing but the usual condition of no 
trapping, and is analogous to the condition on $2M/R$ in the Newtonian case.

Consider next the rate of divergence of the Kretschmann
scalar at the central curvature singularity, as the singularity is
approached in the past along an outgoing null geodesic. Because of the 
dominant energy condition, the leading divergence
in the Kretschmann scalar will be due to the energy density $\rho $. Also, if 
the density is non-singular, this scalar will be finite.

We first transform from the comoving coordinates $(t,r,\theta ,\phi )$ to
coordinates $(t,R,\theta ,\phi )$, thus using the area radius as one of the
coordinates. From equation (\ref{mprime}) it follows that we can write, in
these new coordinates

\begin{equation}
\label{rnew}\rho (t,R)=\frac{(\partial F/\partial R)_t}{R^2}. 
\end{equation}
In order to evaluate the growth of $\rho$ along an outgoing null geodesic,
we use the trajectory of the outgoing null geodesic itself as one of the
coordinates. We label the trajectory by a parameter $X$, which will be a 
function of $t$ and $r$. We eliminate the coordinate $t$ in favour of $X$, and
in the $(X,R,\theta ,\phi )$ coordinates we get the evolution of the density 
to be 
\begin{equation}
\label{cha}\rho (X,R)=\frac{(\partial F/\partial R)_X+(\partial X/\partial
R)_t(\partial F/\partial X)_R}{R^2}. 
\end{equation}

Now, if the density evolution is to be non-singular, $F(X,R)$ must
grow as $R^{3}$ or
slower, as one approaches $R=0$, keeping $X$ fixed. (A faster approach rate 
will cause the density to diverge, as is evident from (\ref{cha})). On the 
other hand, if the 
evolution results in a singularity, we cannot conclude anything about the 
growth rate of $F$ as a function of $R$, at a fixed $X$. This is because the
dominant divergence may come from the second term on the right hand side in
Eqn. (\ref{cha}), rather than the first term. However, we know
from (\ref{thet}) that in order for the singularity to be covered, it is 
necessary that this growth rate be at least as fast as $F(X,R)\sim R$, or 
faster.

We can now give a classification of the nature of the evolution (i.e. whether
it is non-singular, or if singular whether it is naked or covered), depending
on the behaviour of the mass function $F(X,R)$ as a function of $R$, for a 
fixed $X$, in the approach to the singularity. For a given form of matter, 
(say a fluid with a known equation of state, or a massless scalar field)
first consider initial data for which the evolution is non-singular. This 
means that along an outgoing null geodesic, $F$ grows as $R^3$ or slower. 

Consider next changing the initial data so that we come to the class of data 
that lead to singular evolution. We divide such data into two possible 
subclasses, one for which the limiting growth of $F$, for a fixed $X$, is 
$R^{3}$ or slower, and the other for which it is faster than $R^{3}$. The 
singularities in the former class would be naked, since the ratio $F/R$ will 
go to zero in the limit. Hence, if the growth of $F(R)$ is as $R^3$ or 
slower, the evolution is either non-singular or singular and naked.

Those singular evolutions for which the limiting growth rate of $F$ is 
faster than $R^3$ can be further subdivided into three classes. If $F$ grows
faster than $R^3$ but slower than $R$, the singularity is naked. If $F$ grows
as $R$ the singularity is covered if the limiting value of $F/R$ exceeds
unity, and naked if this limiting value is less than or equal to one. If $F$
grows faster than $R$ the singularity will be covered.

In view of the above characterization one can understand better the 
conditions that are required for naked singularity formation in
spherical inhomogeneous dust collapse. For simplicity, we recall the case
of marginally bound dust collapse, for which the initial density profile
near the center, written in a series as 
\begin{equation}
\label{ser}\rho (R)=\rho _0+\rho _1R+\frac 12\rho _2R^2+\frac 16\rho
_3R^3+... 
\end{equation}
determines the outcome of the collapse. It turns out that the singularity is
naked if $\rho _1<0$ or if $\rho _1=0,\rho _2<0$. It is covered if $\rho
_1=\rho _2=\rho _3=0$. If $\rho _1=\rho _2=0$ and $\rho _3<0$ the
singularity is naked if the dimensionless quantity $\zeta =\sqrt{3}\rho
_3/4\rho _0^{5/2}$ is less than or equal to $-25.9904$, and covered when $%
\zeta $ exceeds this value \cite{sj}.

One could ask why the transition from the naked to the covered case takes
place at the level of the third derivative. A physical answer is the
following. It can be shown that in the approach to the central dust
singularity along an outgoing null geodesic, $R\sim r^\alpha ,$ with $\alpha 
$ taking the value $(1+2p/3)$ where $p$ is defined such that the first
non-vanishing derivative in the above expansion for the density is the $p$th
one. According to our naturalness argument above, if the singularity has to
be naked, $F/R^3$ must not vary faster than $1/R^2.$ Noting that $F\sim r^3$%
, we can write $F/R^3\sim R^{3(1-\alpha )/\alpha }$ which implies that $%
\alpha \leq 3$, i.e. $p\leq 3,$ is necessary for nakedness ($p<3$ is
sufficient for nakedness). This physically explains the significance of the
transition taking place at the level of the third derivative. We also note
that the condition that $2R'\zeta^r/R$ does not vanish in the limit is
satisfied in the dust model.

\section{Discussion and Conclusions}

We summarize here the assumptions that we have made: 
(i) We have considered Type I matter fields obeying the dominant energy
condition; (ii) we have assumed 
that the chosen initial conditions are such that any central shell-focusing
singularity that might form in collapse is not preceded by a shell-crossing 
singularity; (iii) in the expression (\ref{thet}) for $\theta$ we have
assumed that the quantity $2R'\zeta^r/R$ remains non-zero in the limit.

Subject to these assumptions, and subject to choosing a specific form of 
matter, we have given a partial classification of 
the nature of evolution (as to whether it is non-singular, and if singular
whether it is covered or naked) in terms of the behaviour of the mass function
in the neighbourhood of the origin. For a given set of initial conditions,
the evolution is non-singular or naked
singular if this growth rate is $R^3$ or slower. If at some epoch during
the evolution the growth rate becomes faster than $R^3$, this shows the
formation of a curvature singularity at that epoch. This singularity is
naked if the growth rate is slower than $R$, it is covered if the growth rate
is faster than $R$, and it could be either covered or naked if the growth rate
is same as $R$. 

We make the following observation about the nature of the shell-focusing
singularity. Suppose it is the case that for a given form of matter, there are 
generic initial data leading to non-singular evolution, and other generic
initial data leading to covered singularities, but the initial data leading
to naked singularities is non-generic. This implies the following regarding
the behaviour of the mass function. There are generic initial conditions
for which the mass function grows as $R^3$ and other generic initial 
conditions for which, near the singularity, it grows as $R$ or faster. 
However, the initial conditions for which it grows faster than $R^3$ but 
slower than $R$ near the singularity are non-generic.

A priori, one cannot draw conclusions as to how the dependence of the mass
function on $R$ changes, when the initial data is changed so as to go from the
non-singular class of solutions to the singular class. If this change is a
continuous one, then there will necessarily be generic data for which $M(R)$ 
grows faster then $R^3$ but slower than $R$, and naked singularity formation 
will be generic. On the other hand, if the change is a discontinuous one, and 
there is a jump from $M(R)\sim R^3$ to $M(R)\sim$ faster than $R$, naked 
singularities will be non-generic (provided also that evolutions of the type 
$M\sim R^3$ are generically non-singular). Whether or not this change is a 
continuous one cannot be decided from the present study, and further 
investigation is necessary.

Although we cannot conclude here about the genericity or non-genericity of 
locally naked
singularities, we compare our results briefly with other recent developments 
relating to naked singularities in spherical gravitational collapse. 
Christodoulou \cite{ch} has shown that for the spherical collapse of a 
massless scalar field, the initial data leading to {\it globally} naked
singularities is non-generic. To our understanding, Christodoulou's result
does not have a bearing on whether or not the {\it locally} naked 
singularities in his model are non-generic. Since the arguments we have given 
above are only for local nakedness, they cannot be directly compared with 
Christodoulou's conclusions. Joshi and Dwivedi \cite{dj} have demonstrated
that the 
formation of locally naked singularities is generic in the spherical collapse 
of Type I matter fields obeying the weak energy condition. Since the weak
energy condition does not necessarily imply the dominant energy condition,
their result does not necessarily imply that $M(R)$ grows faster than $R^3$
and slower than $R$ near the singularity, for generic initial data.

Finally, we comment briefly on numerical studies of spherical collapse. In
these studies (including those on scalar fields and fluids) it has
been found that dispersive evolutions and black hole formation both result from
generic initial data. In many of the the black hole class of solutions, the 
amount of mass that collapses to form the black hole behaves as 
$(p-p_{*})^\gamma$, where $p$ is a parameter labeling a family of solutions,
and $p_{*}$ is the critical value below which the evolution is dispersive.
The limiting solution for which $p\rightarrow p_{*}$ is a naked singularity,
and numerical studies suggest that such a naked singularity is non-generic, 
because it appears to occur only at one value of $p$, i.e. $p=p_{*}$.

To our understanding however, the following important issue should be taken
note of. In most numerical studies to date, the identification of a black
hole in a simulation is carried out via the detection of an apparent horizon.
Such a detection does not by itself rule out the development of a Cauchy 
horizon and a naked singularity which is not covered by the apparent horizon.
Hence, in principle, the solutions that are classified as black holes could
include, as a subset, naked singular solutions. The few simulations that have
been carried out using double null coordinates that can be extended up to the
center of spherical symmetry do not appear to offer a completely clear picture
in this regard. Hence it appears to us that further numerical investigation is
necessary in order to examine the issue of genericity of naked
singularities in spherical collapse.

I acknowledge partial support of
the Junta Nacional de Investigac\~ao Cient\'ifica e Tecnol\'ogica (JNICT)
Portugal, under contract number CERN/S/FAE/1172/97.

\end{document}